\documentclass[%
 aip,
 amsmath,amssymb,
 reprint,%
]{revtex4-1}

\usepackage{graphicx}
\usepackage{dcolumn}
\usepackage{subcaption}
\usepackage{bm}
\usepackage[utf8]{inputenc}
\usepackage[T1]{fontenc}
\usepackage{mathptmx}
\usepackage{etoolbox}

\makeatletter
\def\@email#1#2{%
 \endgroup
 \patchcmd{\titleblock@produce}
  {\frontmatter@RRAPformat}
  {\frontmatter@RRAPformat{\produce@RRAP{*#1\href{mailto:#2}{#2}}}\frontmatter@RRAPformat}
  {}{}
}%
\makeatother
\begin{document}

\preprint{AIP/123-QED}

\title{Lattice anharmonicity effects in fluorite oxide single crystals and anomalous increase in phonon lifetime in ceria at elevated temperature}

\author{A. Khanolkar}
\email[]{*Corresponding author: A. Khanolkar}
\affiliation{Condensed Matter and Materials Physics Group, Idaho National Laboratory, Idaho Falls, ID 83415}

\author{S. Adnan}
\affiliation{Department of Mechanical and Aerospace Engineering, The Ohio State University, Columbus, OH 43210}

\author{M. Minaruzzaman}
\affiliation{Department of Mechanical and Aerospace Engineering, The Ohio State University, Columbus, OH 43210}

\author{L. Malakkal}
\affiliation{Computational Mechanics and Materials Group, Idaho National Laboratory, Idaho Falls, ID 83415}

\author{D.B. Thompson}
\affiliation{Air Force Research Laboratory, Sensors Directorate, Wright-Patterson Air Force Base, Dayton, OH 45433}

\author{D.B. Turner}
\affiliation{Core4ce, Fairborn, Ohio 45324, USA}

\author{J.M. Mann}
\affiliation{Air Force Research Laboratory, Sensors Directorate, Wright-Patterson Air Force Base, Dayton, OH 45433}

\author{D.H. Hurley}
\affiliation{Condensed Matter and Materials Physics Group, Idaho National Laboratory, Idaho Falls, ID 83415}

\author{M. Khafizov}
\email[]{†Corresponding author: M. Khafizov}
\affiliation{Department of Mechanical and Aerospace Engineering, The Ohio State University, Columbus, OH 43210}

\date{\today} 

\begin{abstract}
We investigate the temperature dependence of the frequency and linewidth of the triply-degenerate T$_{2g}$ zone-centered optical phonon in flux-grown ceria and hydrothermally-synthesized thoria single crystals from room temperature to 1273 K using Raman spectroscopy. Both crystals exhibit an expected increase in the phonon linewidth with temperature due to enhanced phonon-phonon scattering. However, ceria displays an anomalous linewidth reduction in the temperature range of 1023–1123 K. First-principles phonon linewidth calculations considering cubic and quartic phonon interactions within temperature-independent phonon dispersion fail to describe this anomaly. A parameterization of the temperature-dependent second order interatomic force constants based on previously reported phonon dispersion measured at room and high temperatures, predicts a deviation from the monotonic linewidth increase, albeit at temperatures lower than those observed experimentally for ceria. The qualitative agreement in the trend of temperature-dependent linewidth suggests that lattice anharmonicity-induced phonon renormalization plays a role in phonon lifetime. Specifically, a change in the overlap between softened acoustic and optical branches in the dispersion curve reduces the available phonon scattering phase space of the Raman active mode at the zone center, leading to an increased phonon lifetime within a narrow temperature interval. These findings provide new insights into higher-order anharmonic interactions in ceria and thoria, motivating further investigations into the role of anharmonicity-induced phonon renormalization on phonon lifetimes at high temperatures.
\end{abstract}

\maketitle
Lanthanide and actinide oxides, such as ceria (CeO$_2$) and thoria (ThO$_2$), are integral to various high temperature technological applications such as catalysis and nuclear power generation that require efficient thermal transport. In these oxides, thermal transport is primarily governed by quantized lattice vibrations or phonons \cite{hurley2022thermal,Zolotarev2021}. At elevated temperatures,the increased vibrational amplitude of atoms in the crystalline lattice, along with the higher occupancy of high-frequency phonons, leads to enhanced phonon-phonon (\textit{ph-ph}) interactions and the renormalization of phonon bands \cite{Leibfried1961}.
Phonon renormalization is evident through changes in the frequencies and lifetimes due to higher-order (or anharmonic) \textit{ph-ph} interactions \cite{Bryan_2020}. The phonon lifetime is determined by phonon interactions, quantified by the scattering phase space as well as third and higher-order interatomic force constants (IFCs). Consequently, these interactions significantly influence phonon-mediated physical properties such as the lattice thermal conductivity \cite{Monserrat2013, Shu2024} and also result in thermal expansion of the lattice. Thus, understanding phonon interactions at higher temperatures is critical for accurately predicting thermophysical properties and the performance of these oxides in energy conversion applications.
Inelastic neutron and X-ray scattering are powerful tools for studying phonon dispersion across entire Brillouin zone. By measuring changes in the energy and momentum of neutrons or X-rays following their interaction with phonons, these techniques provide detailed information about phonon mode frequencies, linewidths, and their interaction strengths \cite{Enda_2022_validating,Ma_2022_APL,Bryan_2020}. On the other hand, Raman spectroscopy is an effective technique for studying the characteristics of optical phonon modes at the Brillouin zone center that satisfy Raman selection rules with much greater accuracy \cite{Lucazeau2003}. This method involves measuring the inelastic scattering of incident photons by the vibrational modes of the crystal. Temperature-dependent measurements of first-order Raman scattering enable the assessment of anharmonic interactions, leading to temperature-dependent shifts (due to phonon softening) and broadening of phonon peaks (due to phonon decay)  \cite{Menendez1984,Balkanski1983,Bauer2009,Chen1995,Dean1982,Gervais1975,Hart1970,Herchen1991,Irmer1996,Mishra2000,Nayar1941,Nelin1974,Pine1969,Su2006,Tang1991,Tsu1982}. These measured shifts and linewidth broadening have been explained by first principles calculations incorporating third-order phonon anharmonicity (i.e., three-phonon scattering processes) \cite{Bechstedt1997,Debernardi1998,Debernardi1995,Debernardi1999,Lang1999,Tang2011} across a range of materials \cite{Lan2012a,Lan2012b,Li2009,Li2011,Morell1996}. Recent density functional theory (DFT) calculations incorporating fourth-order anharmonicity have accurately captured phonon linewidth broadening at high temperatures, a task previously unattainable with only third-order anharmonicity \cite{Yang2020}.
Despite significant interest in lanthanide and actinide oxides for high-temperature applications, studies on anharmonic effects and phonon characteristics at elevated temperatures in these materials are limited. Existing research includes temperature-dependent studies on the T\textsubscript{2g} Raman-active mode in single crystal ceria up to 800 K \cite{Sato1982} and in nanocrystalline powdered ceria up to 1373 K \cite{DograPandey2015,Popovic2007}. While the temperature dependence of the Raman linewidth in single crystal ceria can be described by the cubic anharmonic terms up to 800 K \cite{Sato1982}, four-phonon processes dominate in ceria nanocrystals \cite{Popovic2007} at higher temperatures. Differences in the phonon linewidths between single crystal, polycrystalline, and nanocrystalline ceria have been attributed to phonon confinement effects \cite{Popovic2007, Spanier2001, Weber1993}. Studies on actinide oxides include those on polycrystalline urania pellets, examining the temperature dependence of the T\textsubscript{2g} \cite{Elorrieta2018,Guimbretiere2015} and 2LO modes \cite{Elorrieta2018} up to 873 K. To our knowledge, similar studies on thoria have not been reported. While most studies have provided insights into phonon anharmonicity in polycrystalline actinide oxides, investigations on single crystals, necessary to elucidate intrinsic vibrational properties without extrinsic factors like grain boundary scattering, are scarce primarily due to the challenges in growing high-quality single crystals with acceptable purity and stoichiometry \cite{Mann2010,wanklyn1984flux}.

\begin{figure*}[!htb]
    \centering
    \includegraphics[width=0.8\textwidth]{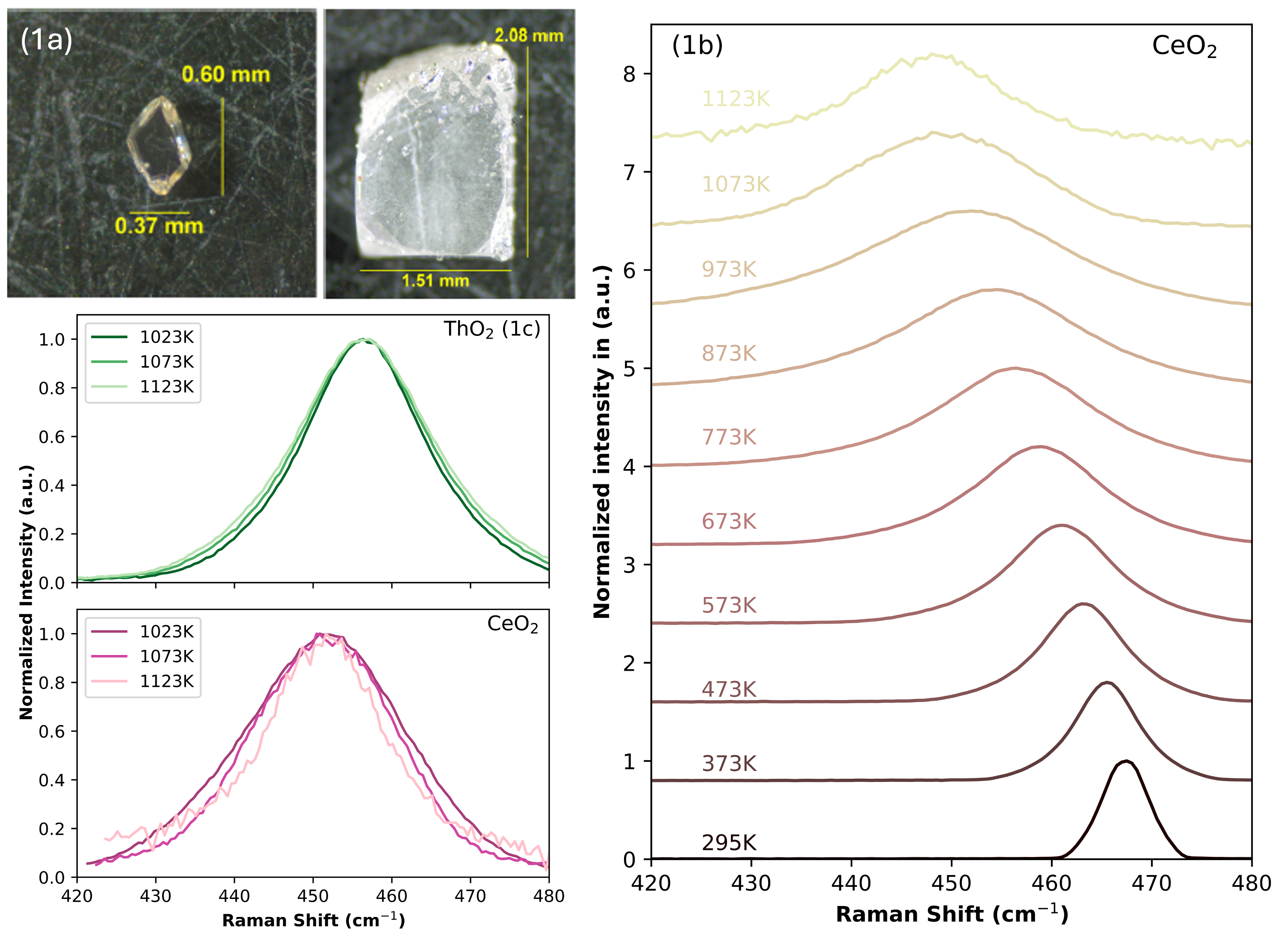}
    \caption{(a) Top-down optical images of ceria and thoria crystals. (b) Temperature-dependent frequency shifts and broadening of the T\textsubscript{2g} Raman-active mode in ceria. (c) Normalized T\textsubscript{2g} peaks of thoria (top) and ceria (bottom) in the 1023–1123 K range, with center frequencies shifted to a fixed value to emphasize linewidth changes with temperature.}
    \label{fig:1}
\end{figure*}

In this study, we address this knowledge gap by investigating anharmonicity-induced changes to phonon characteristics up to 1273 K in ceria and thoria single crystals. Single crystals of ceria and thoria, as shown in Figure~\ref{fig:1} (a), were grown using the flux growth \cite{wanklyn1984flux} and hydrothermal synthesis techniques \cite{Mann2010}, respectively. Synthesis details are previously described in the Ref. \cite{Mann2010}. \textit{In situ} Raman measurements were performed with 532 nm optical excitation on the \{001\} facet of the ceria and thoria crystals. Single crystal were mounted inside a Linkam\textsuperscript{\textregistered} optical heating stage. Raman spectra were collected during the heating and cooling at each temperature step. Experimental setup details are provided in section S1 of the Supplementary Material \cite{material44}. Figure~\ref{fig:1}(b) shows Raman spectra for ceria, highlighting the T\textsubscript{2g} mode's frequency reduction and broadening with temperature. Figure~\ref{fig:1}(c) shows normalized spectra from 1023–1123 K for both crystals. At room temperature, the T\textsubscript{2g} peaks were approximately 465 cm\textsuperscript{-1} for ceria and 464 cm\textsuperscript{-1} for thoria, in agreement with previous reports on flux-grown crystals \cite{keramidas1973raman}. The T\textsubscript{2g} peak for ceria redshifted and broadened with temperature, as shown in Figure~\ref{fig:1}(b).

\begin{figure*}[!htb]
    \centering
    \begin{subfigure}[b]{0.45\textwidth}
        \includegraphics[width=\textwidth]{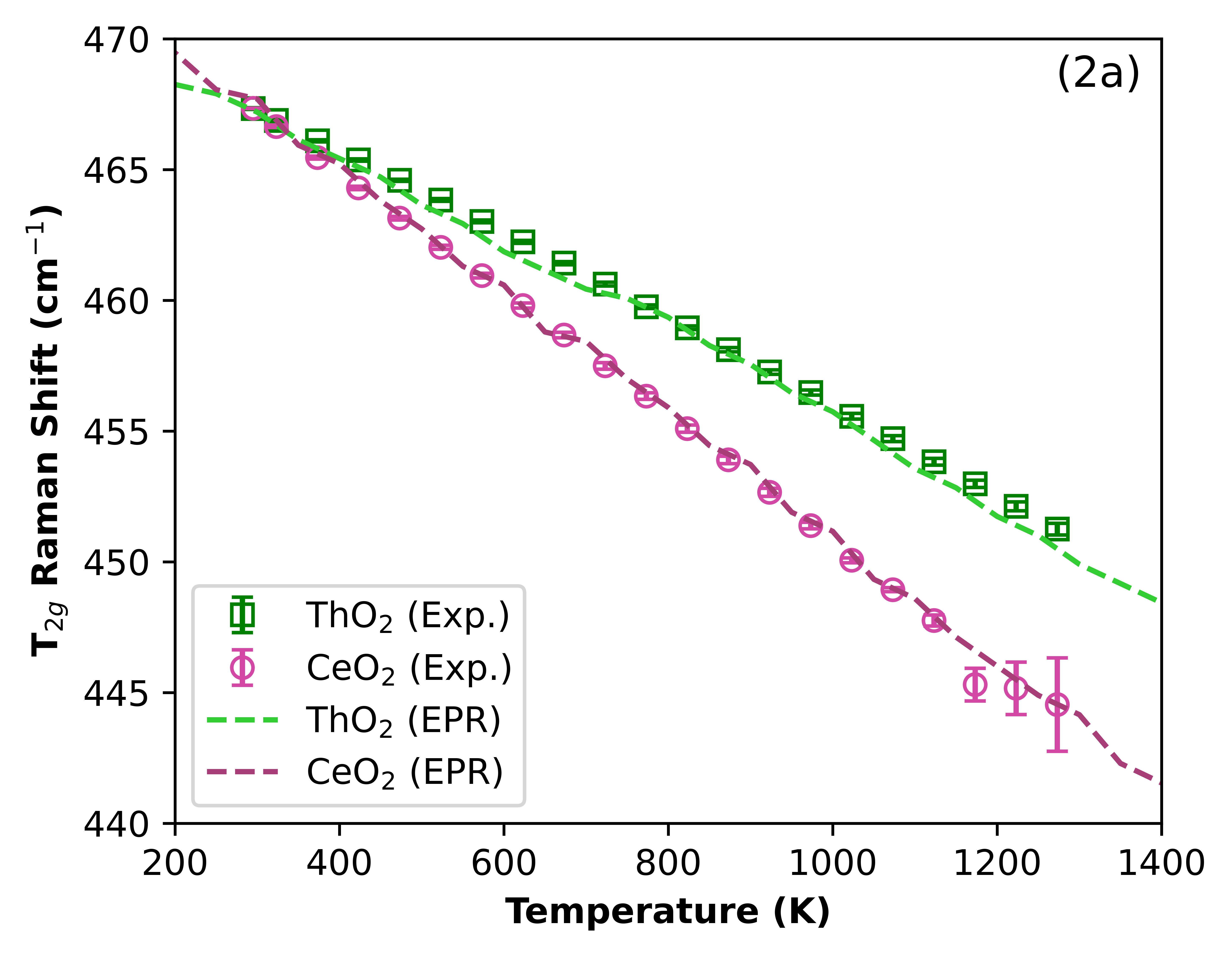}
    \end{subfigure}
    \begin{subfigure}[b]{0.45\textwidth}
        \includegraphics[width=\textwidth]{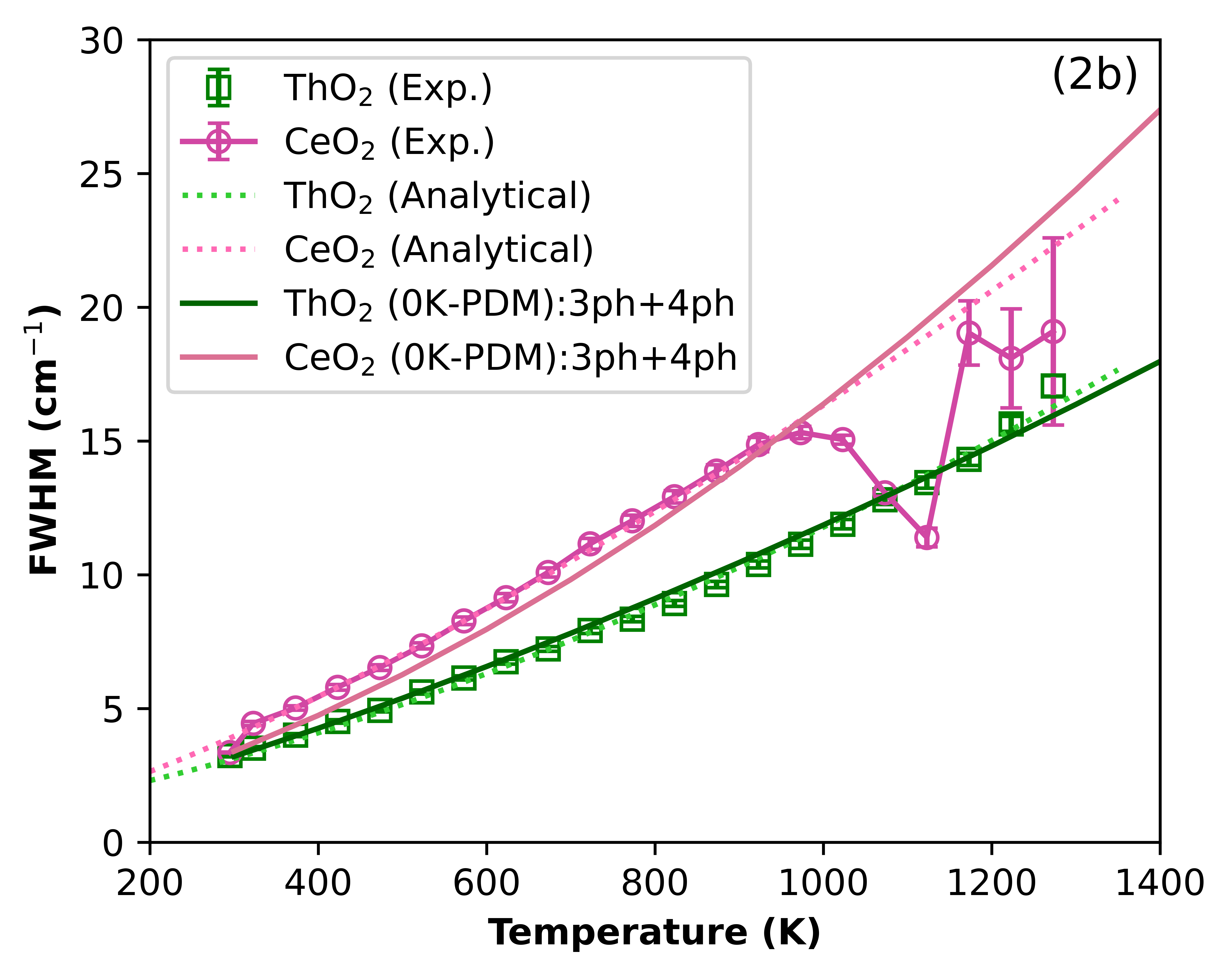}
    \end{subfigure}
    \begin{subfigure}[b]{0.45\textwidth}
        \includegraphics[width=\textwidth]{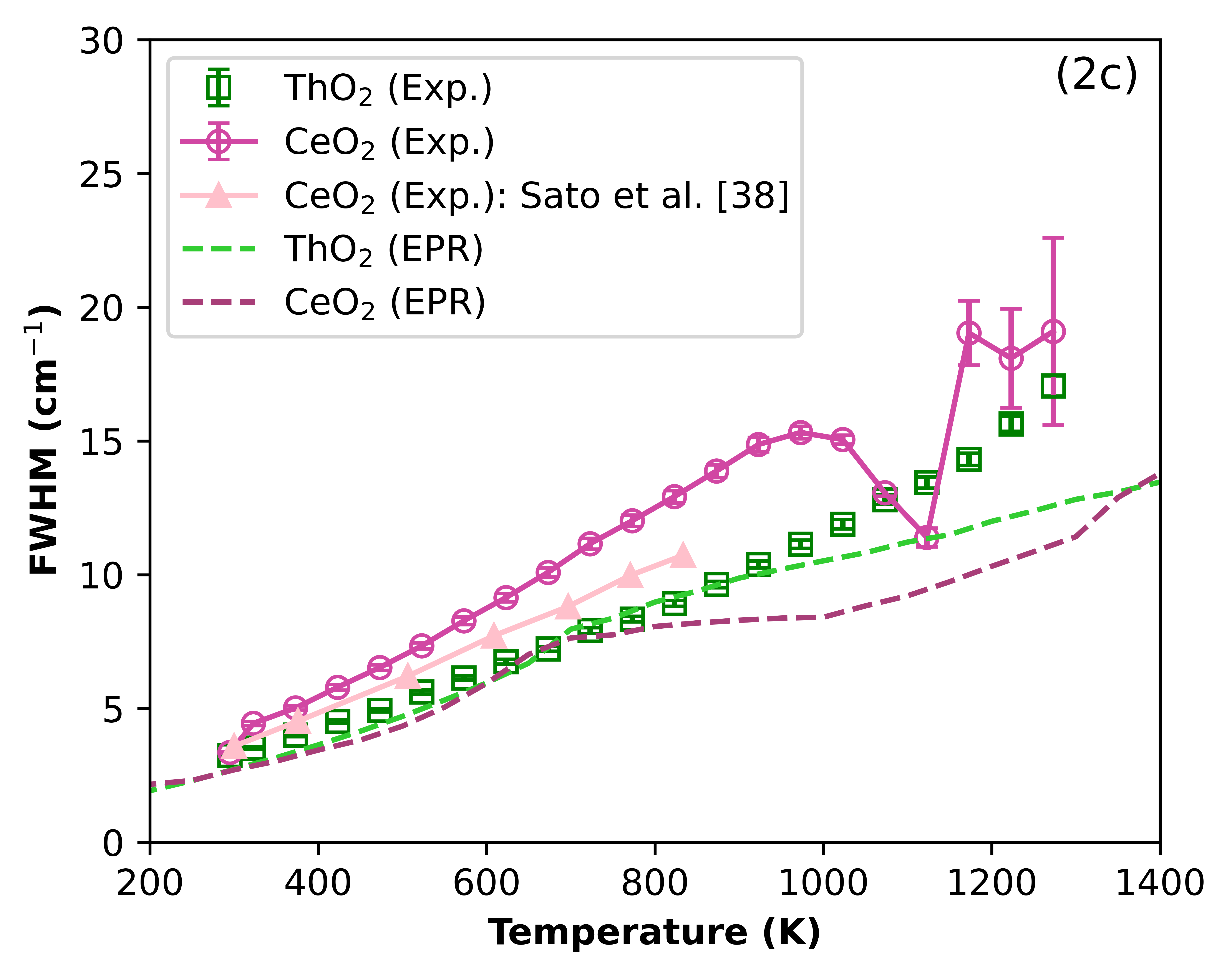}
    \end{subfigure}
    \begin{subfigure}[b]{0.45\textwidth}
        \includegraphics[width=\textwidth]{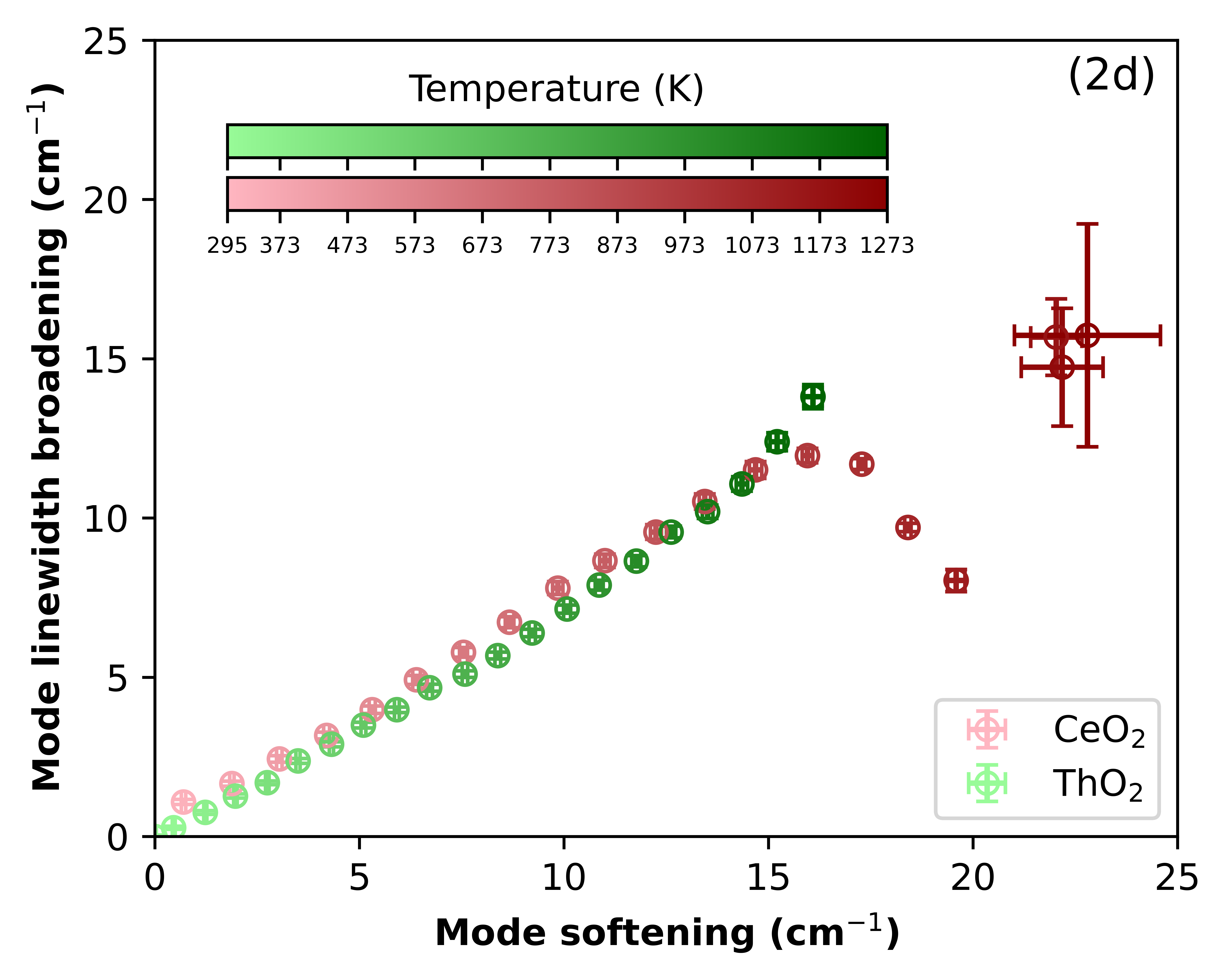}
    \end{subfigure}
    
    \caption{(a) Temperature-dependence of the Raman shift of the T\textsubscript{2g} mode in ceria and thoria (indicated by markers). Error bars represent the 95\% confidence interval of the peak position from the Gaussian fit. The dashed lines denote values obtained from the empirical model with phonon renormalization (EPR). (b) Measured (markers) and calculated changes in the linewidth of the T\textsubscript{2g} vibrational mode with temperature for ceria and thoria using an analytical model fit (dotted lines) and the Zero-Kelvin phonon dispersion method (0K-PDM) without phonon renormalization (solid lines). (c) Comparison of the measured linewidth broadening in ceria and thoria with a previous report on ceria and calculated linewidths using the empirical model fit with phonon renormalization (EPR) . (d) Measured broadening versus softening of the T\textsubscript{2g} vibrational mode in ceria and thoria.}
    \label{fig:2}
\end{figure*}

The center frequency of the T\textsubscript{2g} peak as a function of temperature is shown in Figure~\ref{fig:2}(a). The T\textsubscript{2g} peak in thoria shifts monotonically from $\sim$464 cm\textsuperscript{-1} at room temperature to $\sim$451 cm\textsuperscript{-1} at 1273 K. In contrast, the T\textsubscript{2g} peak in ceria shifts more significantly from $\sim$465 cm\textsuperscript{-1} at room temperature to $\sim$444 cm\textsuperscript{-1} at 1273 K. Figure~\ref{fig:2}(b) shows the temperature dependence of the phonon linewidth of the T\textsubscript{2g} mode in both materials. Both exhibit an increase in linewidth up to $\sim$1000 K due to increased phonon-phonon scattering and reduced phonon lifetime. However, while the thoria's linewidth continues to broaden beyond 1000 K, ceria shows an anomaly in the linewidth of the T\textsubscript{2g} between 1023–1123 K. This opposite and counter-intuitive trend in linewidth is evident in the normalized spectra in Figure~\ref{fig:1}(c). Beyond 1123 K, the ceria's phonon linewidth decreases abruptly and then gradually increases with temperature. The larger uncertainty in the phonon linewidth measured above 1200 K is attributed to the reduced signal-to-noise ratio of the T\textsubscript{2g} peak in the Raman spectra acquired at elevated temperatures. Figure~\ref{fig:2}(d) illustrates the relationship between T\textsubscript{2g} peak broadening and mode softening in both materials. Despite greater softening and broadening in ceria, the relationship is nearly identical for both. The anomaly in the tmemperature-dependent linewidth of ceria suggests a mechanism reducing phonon scattering in 1023–1273 K range. Comparing ceria's linewidth broadening with Sato and Tateyama's data \cite{Sato1982} (Figure~\ref{fig:2}(c)), we see reasonable agreement at 300 K and 350 K, but a clear deviation at higher temperatures, likely due to differences in sample quality and purity. To the best of our knowledge, similar measurements of phonon softening and linewidth broadening at elevated temperature have not been previously reported for thoria.    

To elucidate the role of anharmonicity in Raman peak shifts and linewidth changes, we use three lattice dynamics-based modeling approaches. The first approach, herein referred to as the `analytical model', uses an expression developed by Klemens \cite{klemens1966anharmonic}. This model considers the decay of an optical phonon into two acoustic phonons (i.e., 3-phonon scattering) at the $\Gamma$ point and extends it to include 4-phonon processes for explaining the temperature dependence of the phonon linewidth \cite{Balkanski1983}. It is based on the Debye density of states and averaged IFCs, describing the temperature-dependence of the phonon linewidth broadening $\Gamma(T)$ as,

\begin{multline}
\Gamma(T) = \Gamma(0) + A \left[ 1 + \frac{2}{e^{\frac{\hbar \omega(0)}{2k_B T}} - 1} \right] \\
+ B \left[ 1 + \frac{3}{e^{\frac{\hbar \omega(0)}{3k_B T}} - 1} + \frac{3}{\left(e^{\frac{\hbar \omega(0)}{3k_B T}} - 1\right)^2} \right],
\end{multline}

where $\omega = 464$ cm\textsuperscript{-1} is the Raman shift at 0 K, calculated using the rigid-ion \cite{Weber1993} or shell model \cite{hurrell1970crystal}, $k_B$ is the Boltzmann constant and $T$ is the temperature. Parameters $A$ and $B$ represent contributions from three-phonon and four-phonon processes, respectively. Fitting the measured linewidth to the Equation (1) gives $A = 1.21$ cm\textsuperscript{-1} and $B = 0.0675$ cm\textsuperscript{-1} for thoria, and $A = 2.19$ cm\textsuperscript{-1} and $B = 0.0629$ cm\textsuperscript{-1} for ceria. These fits are shown by dotted lines in Figure~\ref{fig:2}(b). Note that data up to $\sim$920 K were used for the ceria fit, as the analytical model does not capture the anomalous linewidth reduction observed in our experiment. The higher values of the $A$ parameter compared to $B$, suggest that linewidth broadening is dominated by 3-phonon processes, though there is a significant contribution from 4-phonon processes. Setting the $B$ parameter to zero, thus only considering A, fails to accurately capture the temperature-dependent linewidth trends of thoria and ceria. This indicates that while 4-phonon processes play a smaller role, including their contribution is necessary to describe the linewidth broadening of the T\textsubscript{2g} phonon. Additionally, the higher $A$ value for ceria compared to thoria suggests more significant 3-phonon scattering in ceria at the same temperature, which aligns with the observed greater phonon softening and linewidth broadening observed in ceria. 

In the second modeling approach, we determined the phonon linewidths of thoria and ceria entirely from first-principles calculations combined with the Boltzmann transport equation (BTE) formalism \cite{maradudin1962thermal}. We refer to this as the `zero-kelvin phonon dispersion method (0K-PDM)'. Using the Phonopy code \cite{phonopy_JPCM,phonopy_JPSJ}, we computed the second-order force constants at the ground state, neglecting anharmonicity-induced phonon renormalization and thus not considering the Raman peak shift with temperature. The linewidth contribution due to the third-order \cite{ShengBTE_3ph} and fourth-order force \cite{ShengBTE_4ph} constants were computed using the ShengBTE code \cite{shengBTE_2014}. Details of the first-principles calculations are provided in Section S2 of the Supplemental Material \cite{material44}. Figure~\ref{fig:2}(b) shows the linewidths predicted by the 0K-PDM approach. The solid lines indicate the temperature-induced changes in phonon linewidth for both thoria and ceria. The predicted linewidths for thoria match the experimental measurements well, while ceria predictions accurately capture the linewidth broadening up to around 1000 K. However, the approach fails to explain the reduction in phonon linewidth observed in ceria between 1023 K and 1123 K. Although the 0K-PDM considered three- and four-phonon interactions, it neglected effects of anharmonic phonon renormalization which may contribute to changes in scattering phases space. \cite{errea2014anharmonic,tadano2015self,xia2018revisiting,xia2020high}. 

Given that both the analytical and the 0K-PDM models failed to capture the anomalous reduction, an alternative explanation is needed. A recent experimental study reported a similar anomalous phonon linewidth sharpening at elevated temperatures for the longitudinal acoustic mode of lead selenide (PbSe) at 770 K using inelastic X-ray scattering \cite{manley2019intrinsic}. This was attributed to a shrinking of the available phase space for scattering with increasing temperature. We hypothesize that the anomaly in the T\textsubscript{2g} mode lifetime in ceria at elevated temperatures is due to a similar effects, where lattice anharmonicity-induced phonon renormalization. Therefore, our third modeling approach, the empirical model with phonon renormalization (EPR), aims to capture phonon renormalization effects using an empirical method based on previously reported experimental dispersion curves of ceria and thoria \cite{clausen1987inelastic,mathis2022generalized}.

In the EPR model, we account for the temperature dependence of the second-order IFCs while considering only three-phonon interactions and neglecting changes in the third-order IFCs. The second-order IFCs of thoria were fitted to previously reported temperature-dependent phonon dispersion curves by treating phonon interactions traditionally considered by rigid ion models \cite{Lacina_PRB_1970,Nakajima_PRB_1994} . The phonon dispersion curves used for these fits were obtained from measurements on thoria at 5 K, 300 K, and 750 K using inelastic neutron scattering \cite{mathis2022generalized}. The temperature dependent second-order IFCs in thoria were obtained by interpolating the experimental phonon dispersion data using the Lakkad expression (based on the Debye model of phonon dispersion) \cite{lakkad1971temperature},

\begin{equation}
F_{\alpha \beta} = F_{\alpha \beta}^0 [1-K_{\alpha \beta} F(T/\theta_D)],
\end{equation}
where $F$ is 
\begin{equation}
F(T/\theta_D) = 3\left(\frac{T}{\theta_D}\right)^4 \int_0^{\theta_D/T} x^3 \left[\exp(x)-1\right] dx,
\end{equation}

$\theta_D$ is the Debye temperature and $K_{\alpha \beta}$ is the anharmonicity parameter for each IFC \cite{Leibfried1961,khanolkar2023temperature}. Third-order IFCs of thoria, at 0 K were obtained from previously reported first-principle calculations \cite{jin2021assessment}. The phonon linewidth as a function of temperature was then extracted using the Phono3py package with a q-point mesh of 20 x 20 x 20.  \cite{maradudin1962thermal,togo2023first,togo2015distributions}. For ceria, due to the availability of only room temperature phonon dispersion data \cite{clausen1987inelastic}, the Lakkad model parameters could not be uniquely defined. Therefore, a set of parameters that reproduced the room temperature dispersion was chosen. The temperature-dependent phonon linewidth of ceria was then calculated using the 2nd order IFCs obtained using this approach. Parameters of the Lakkad model parameters and thermally correlated force constants are provided in Section S2 of the Supplementary Material \cite{material44}.
\begin{figure*}[!htb]
    \centering
    \includegraphics[width=0.8\textwidth]{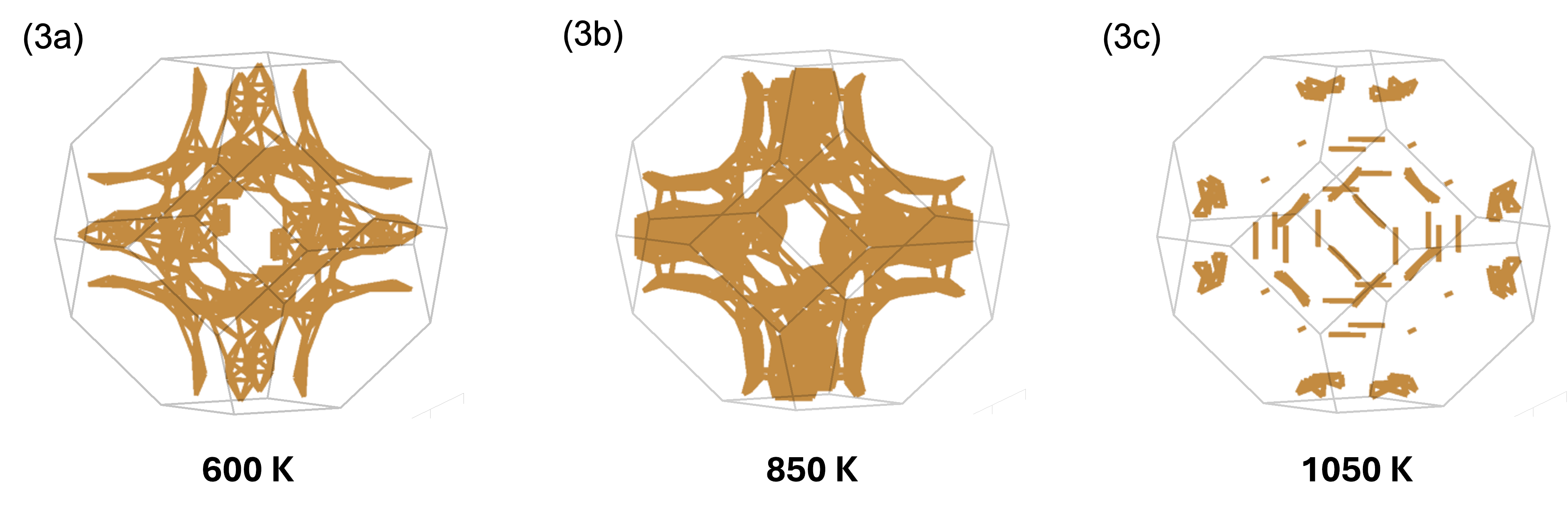}
    \caption{Temperature-dependent phase space for 3-phonon scattering processes in ceria at (a) 600 K, (b) 850 K, and (c) 1050 K. These surfaces in the Brillouin zone represent scattering processes allowed by energy and momentum conservation for the T\textsubscript{2g} mode fixed at the zone center, $\omega_1 = T\textsubscript{2g}(\Gamma)$, the transverse acoustic (TA) phonon, $\omega_2 = TA(q_2)$, and the longitudinal optic phonon (LO), $\omega_3 = LO(-q_1 - q_2)$. The orange surface represents allowed values for $q_2$.}
    \label{fig:fig3}
\end{figure*}
Using the EPR model, we first predict the frequency shift of the T\textsubscript{2g} mode and compare it with the experimental measurements, as shown in Figure~\ref{fig:2}(a). The agreement between the measured frequency shift of the T\textsubscript{2g} mode and the EPR model calculations is excellent. Furthermore, this approach successfully captures the linewidth broadening at low and moderate temperatures (up to approximately 900 K) for thoria, as shown in Figure~\ref{fig:2}(c). However, for ceria, the linewidths calculated using the EPR model consistently underestimate the measured values by $\sim$\textit{30-35$\%$}, although the general trend with temperature is similar up to around 700 K. Beyond this temperature, a deviation between the calculated and measured phonon linewidths is observed. This is expected since higher-order phonon scattering processes become significant at elevated temperatures, while the calculations consider only three-phonon interactions and neglect higher-order mechanisms. Notably, the calculated linewidths exhibit a non-monotonic increase with temperature, attributed to changes in the three-phonon scattering phase space as the phonon dispersion curves are renormalized. This is in contrast to the 0K-PDM without phonon renormalization, which shows a monotonic increase in the T\textsubscript{2g} mode linewidths with temperature for both materials, as depicted in Figure~\ref{fig:2}(b).

From Figure~\ref{fig:2}(c), we observe that while the temperature-dependent trends in the linewidths obtained from the EPR model predict an anomalous narrowing of the phonon linewidth at a lower temperature than observed experimentally in ceria, they still provide a qualitative explanation for the possible mechanism behind the observed anomaly. In the case of thoria, our analysis suggests that the anomalous linewidth narrowing occurs at an elevated temperature outside our measurement range. To further elucidate the origin of the reduction in the phonon scattering phase space resulting from anharmonic renormalization of the phonon band at elevated temperatures, we plot the scattering phase space in ceria at different temperatures using the temperature-dependent IFCs obtained from the EPR model, as shown in Figure~\ref{fig:fig3}. The orange surfaces in the Brillouin zone depicted in Figure~\ref{fig:fig3} represent the three-phonon scattering processes allowed by energy and momentum conservation of the T\textsubscript{2g} mode at the zone center. A clear reduction in the surface area representing the phonon scattering phase space is seen at 1050 K. This observation further corroborates the role of phonon renormalization in narrowing the phonon lifetime at elevated temperatures.

Zone-centered optical phonons generally contribute minimally to the thermal conductivity of fluorite oxides due to their low group velocity and short relaxation times \cite{li2021contribution}. First-principles calculations indicate that the triply-degenerate T\textsubscript{2g} mode contributes approximately 14.5\% to the lattice thermal conductivity of thoria \cite{malakkal2019thermal} at room temperature and approximately 13.7\% in ceria \cite{malakkal2019atomistic}. However, in conditions such as increased structural complexity \cite{li2021contribution} or strong overlap between acoustic and optical phonon branches away from the Brillouin zone center \cite{zhang2012first}, optical phonons significantly influence the lattice thermal conductivity of crystalline materials. Despite the minimal contribution of the T\textsubscript{2g} mode in ceria and thoria to thermal conductivity \cite{malakkal2019thermal,malakkal2019atomistic,zhou2022improving}, renormalization of acoustic and optical phonon branches at elevated temperatures can notably affect lattice thermal conductivity. It is crucial to consider that increased phonon lifetimes due to reduced scattering from phonon renormalization may enhance the lattice thermal conductivity of ceria and thoria at higher temperatures \cite{Yang_2022}. This is particularly relevant for applications such as next-generation nuclear fuels or solid-state electrolytes, where operating temperatures fall within the range where phonon linewidth narrowing and lifetime enhancement occur. Further investigation into the high-temperature phonon dispersion and lattice thermal conductivity of ceria and thoria is necessary to definitively establish the impact of phonon linewidth narrowing on thermal conductivity.

In summary, we have reported the temperature-dependent mode softening and linewidth broadening of the zone-centered T\textsubscript{2g} mode in hydrothermally synthesized ceria and thoria single crystals in the temperature range of 300–1273 K using Raman spectroscopy. Our findings reveal that the reduction in phonon frequency and increase in phonon lifetime are more pronounced in ceria compared to thoria. By integrating the temperature-dependent softening data with previously reported volume expansion  data, we derived the mode-specific Grüneisen parameters for both materials. Notably, T\textsubscript{2g} phonon linewidth generally increased with temperature, but an anomalous reduction was observed in ceria between 1023–1123 K. While DFT calculations with ground-state interatomic force constants and 4th order phonon interactions failed to capture this anomaly, calculations incorporating temperature-dependent second-order force constants and 3-phonon interactions qualitatively matched the experimental observations at a slightly lower temperature. However, calculations that accounted for the temperature dependence of the second-order force constants with 3-phonon interactions qualitatively captured the anomaly but at a lower temperature than observed experimentally. We attribute this anomaly to anharmonic phonon renormalization at elevated temperatures, which reduces the effective phonon scattering phase space at the Brillouin zone center. Future work should focus on higher-order scattering events and detailed phonon dispersion measurements at elevated temperatures to achieve precise descriptions of phonon dynamics in these fluorite oxides. These insights will be crucial for evaluating the lattice thermal conductivity and performance of these materials in applications such as next-generation nuclear fuels, high-temperature catalysts, and solid-state fuel cell electrolytes. Our study provides a foundation for further investigation into the impact of phonon renormalization on the thermal properties of fluorite oxides, advancing our understanding of their potential in advanced technological applications.

This work was supported by the Center for Thermal Energy Transport under Irradiation (TETI), an Energy Frontier Research Center funded by the U.S. Department of Energy, Office of Science, Office of Basic Energy Sciences.The authors also acknowledge that this research made use of the resources of the High Performance Computing Center at Idaho National Laboratory, which is supported by the Office of Nuclear Energy of the U.S. Department of Energy and the Nuclear Science User Facilities under Contract No. DE-AC07-05ID14517. This manuscript has been authored by Battelle Energy Alliance, LLC under Contract No. DE-AC07-05ID14517 with the U.S. Department of Energy. The United States Government retains and the publisher, by accepting the article for publication, acknowledges that the U.S. Government retains a nonexclusive, paid-up, irrevocable, world-wide license to publish or reproduce the published form of this manuscript, or allow others to do so, for U.S. Government purposes.

\appendix
\section{Supplementary Information}
\section*{S1. Experimental Details}

\textit{In situ} Raman measurements were performed by placing the crystals in a Linkam\textsuperscript{\textregistered} TS1000 temperature-controlled stage, coupled to an Olympus\textsuperscript{\textregistered} MPlan N 10× objective lens (numerical aperture of 0.25) connected to the turret of a Horiba Jobin Yvon\textsuperscript{\textregistered} LabRAM HR800 confocal Raman microscope. Raman excitation was provided by a Coherent\textsuperscript{\textregistered} Verdi 532 nm continuous wave (CW) laser, with the beam focused on the \{001\} facet of the ceria and thoria crystals through the silica window of the heating stage, with an incident power of \(\sim 450 \ \mu\text{W}\). Raman spectra were acquired during the heat-up and cool-down phases of the experiment at each temperature step. Figure \ref{fig:supp_fig1} shows an image of the ceria crystal in the Linkam heating stage at 1273 K during a Raman spectrum acquisition.
\renewcommand{\thefigure}{S\arabic{figure}}
\setcounter{figure}{0}

\begin{figure}[h]
    \centering
    \includegraphics[width=0.4\textwidth]{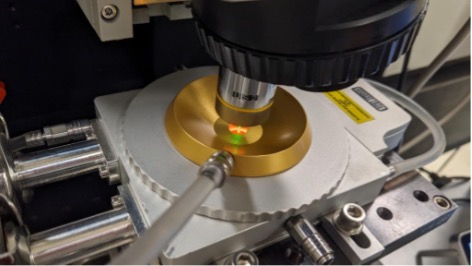}
    \caption{Experimental setup showing the ceria crystal in the Linkam heating stage during a Raman spectrum acquisition.}
    \label{fig:supp_fig1}
\end{figure}

\section*{S2. Details of the theoretical models}
\subsubsection{0K-PDM}

All density functional theory \cite{Kohn_1965} calculations in this study were performed using the projector-augmented-wave \cite{blochl_1994} method within the Vienna Ab initio Simulation Package (VASP) \cite{Kresse_1996}, employing the local density approximation (LDA) \cite{Perdew_1981} for exchange and correlation. Geometry optimization of ThO$_{2}$ and CeO$_{2}$ (space group $Fm\bar{3}m$) was performed on a primitive unit cell by minimizing total energy with respect to cell parameters and atom positions using the conjugate gradient method. Energy convergence for these materials was achieved with an electron wave vector grid of 12$\times$12$\times$12  and a plane wave energy cutoff of 550 eV, with electronic energy convergence criteria set at $10^{-8}$ eV. Given the polar nature of both CeO$_{2}$ and ThO$_{2}$, non-analytical contributions were considered. The Born charges and dielectric constants required for evaluating the non-analytical corrections were calculated using density functional perturbation theory \cite{Baroni_2001_Phonons}. For ThO$_{2}$, harmonic force constants were evaluated using a 5 × 5 × 5 supercell with 375 atoms and a k-point grid of 2$\times$2$\times$2  via the finite displacement method implemented in PHONOPY \cite{TOGO2015}. Third-order (anharmonic) force constants were calculated using a 5$\times$5$\times$5  supercell at the gamma point with Thirdorder.py \cite{shengBTE_2014}, setting the force cutoff distance to the fifth nearest neighbors. Fourth-order force constants were determined using a 4$\times$4$\times$4  supercell at the gamma point with Fourthorder.py \cite{ShengBTE_4ph}, setting the force cutoff distance to the second nearest neighbors. For CeO$_{2}$, second order force constant was calculated on a 5$\times$5$\times$5 supercell and k-points of 3$\times$3$\times$3 , the third-order force constants were calculated using a  5$\times$5$\times$5  supercell, setting the force cutoff distance to the third nearest neighboring atoms. For the fourth-order force constant we used a 4$\times$4$\times$4  supercell at the gamma point, setting the force cutoff distance to the second nearest neighboring atoms. The lattice thermal conductivity (\textit{$k_{L}$}) was calculated using the iterative solutions of the Boltzmann transport equation (BTE) as implemented in ShengBTE \cite{shengBTE_2014}. The phonon linewidth was calculated from scattering rates. Further details of the calculations of ThO$_{2}$  work are available in reference \cite{Malakkal_PRM}. 

\begin{figure*}[ht]
    \centering
    \hspace{0.8\textwidth}
    \begin{subfigure}{0.4\textwidth}
        \includegraphics[width=\textwidth]{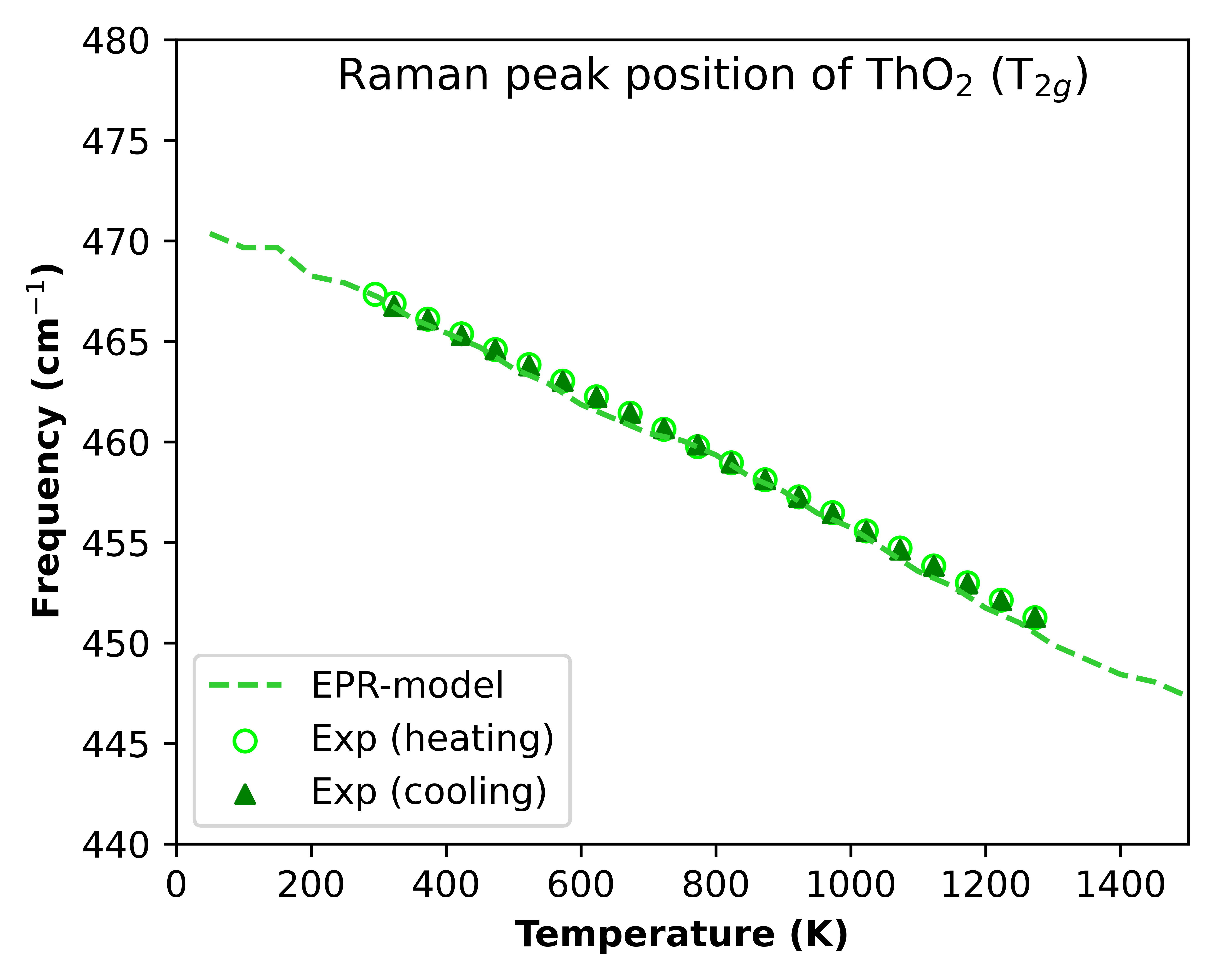}
        \caption{}
        \label{fig:Figure_S2a}
    \end{subfigure}
    \hspace{0.01\textwidth}
    \begin{subfigure}{0.4\textwidth}
        \includegraphics[width=\textwidth]{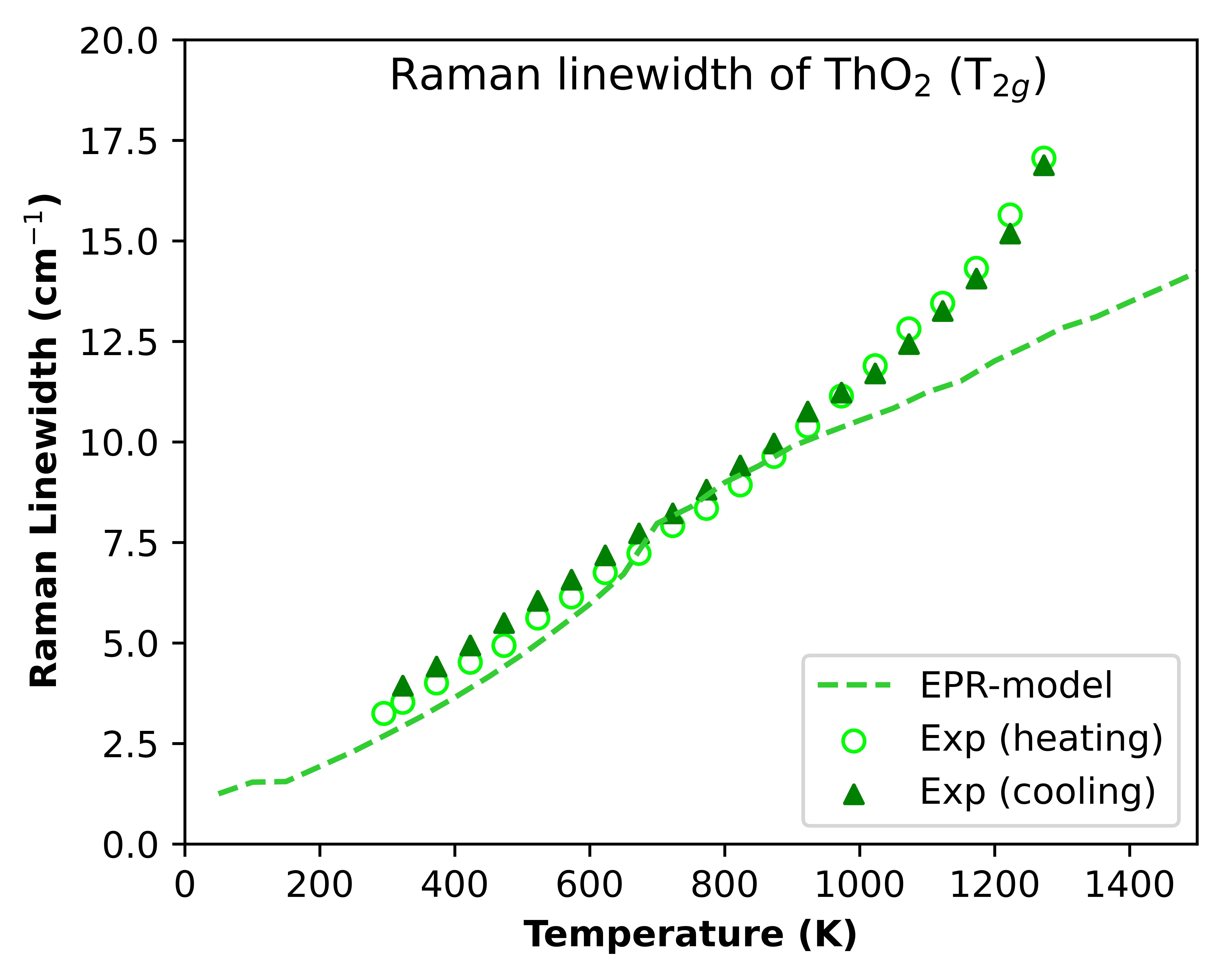}
        \caption{}
        \label{fig:Figure_S2b}
    \end{subfigure}
    \hspace{0.01\textwidth}
    \caption{Temperature-dependence of the (a) frequency and (b) linewidth of the T$_{2g}$ mode in ThO$_{2}$ obtained from the EPR model}
    \label{fig:Figure_S2}
\end{figure*}

\begin{figure*}[ht]
    \centering
    \hspace{0.8\textwidth}
    \begin{subfigure}{0.4\textwidth}
        \includegraphics[width=\textwidth]{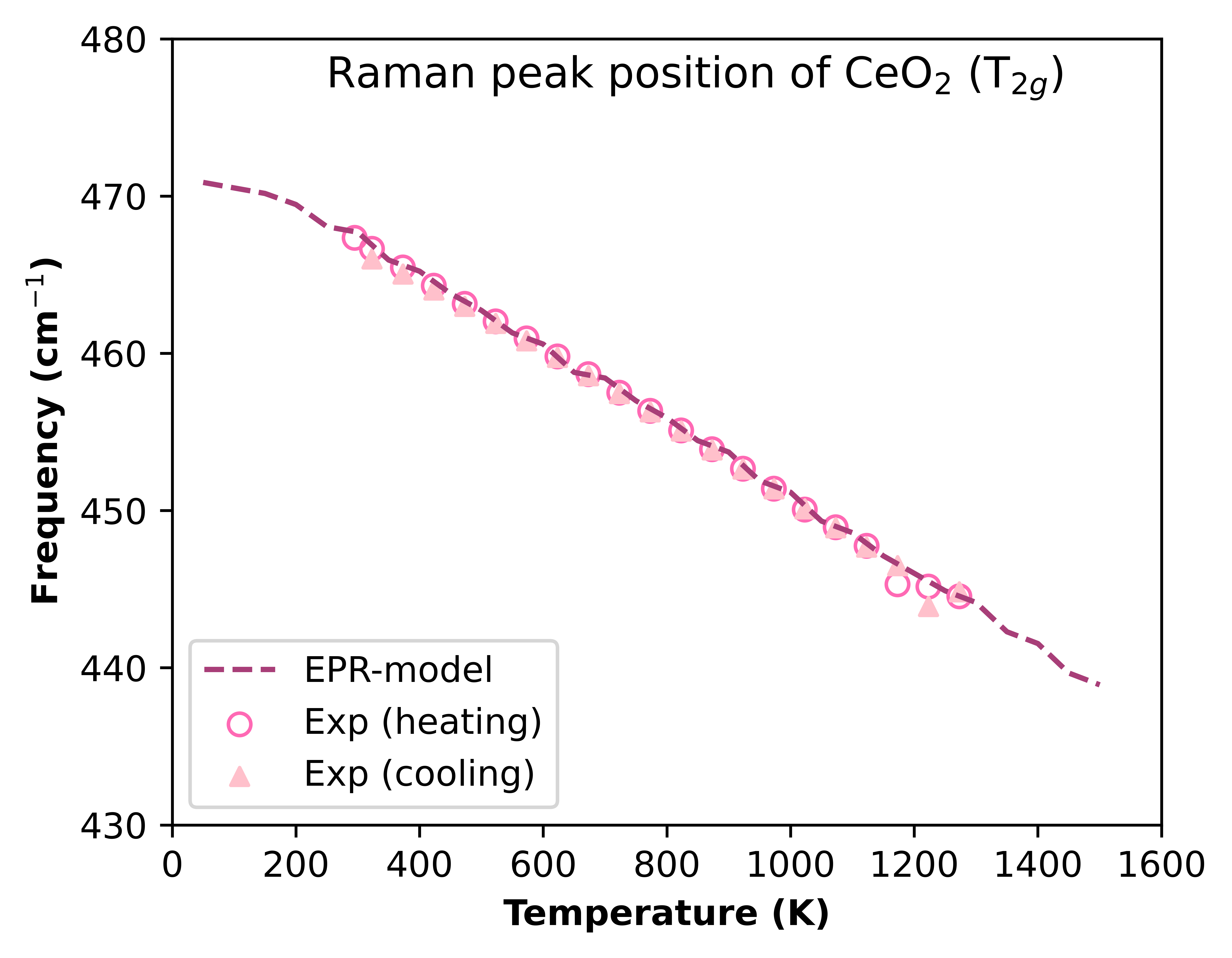}
        \caption{}
        \label{fig:Figure_S3a}
    \end{subfigure}
    \hspace{0.01\textwidth}
    \begin{subfigure}{0.4\textwidth}
        \includegraphics[width=\textwidth]{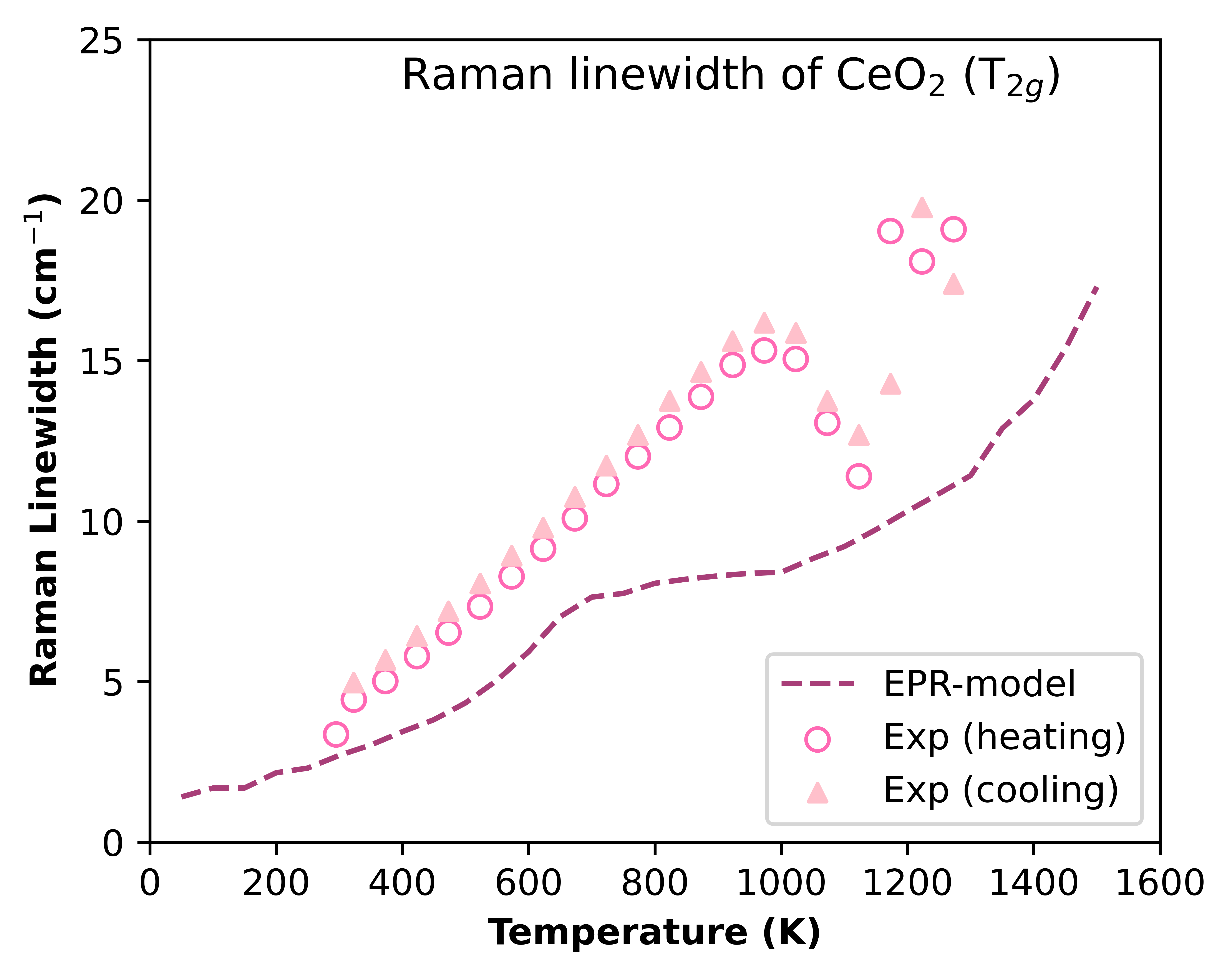}
        \caption{}
        \label{fig:Figure_S3b}
    \end{subfigure}
    \hspace{0.01\textwidth}
    \caption{Temperature-dependence of the (a) frequency and (b) linewidth of the T$_{2g}$ mode in CeO$_{2}$ obtained from the EPR model}
    \label{fig:Figure_S3}
\end{figure*}

\begin{figure*}[ht]
    \centering
    \hspace{0.8\textwidth}
    \begin{subfigure}{0.4\textwidth}
        \includegraphics[width=\textwidth]{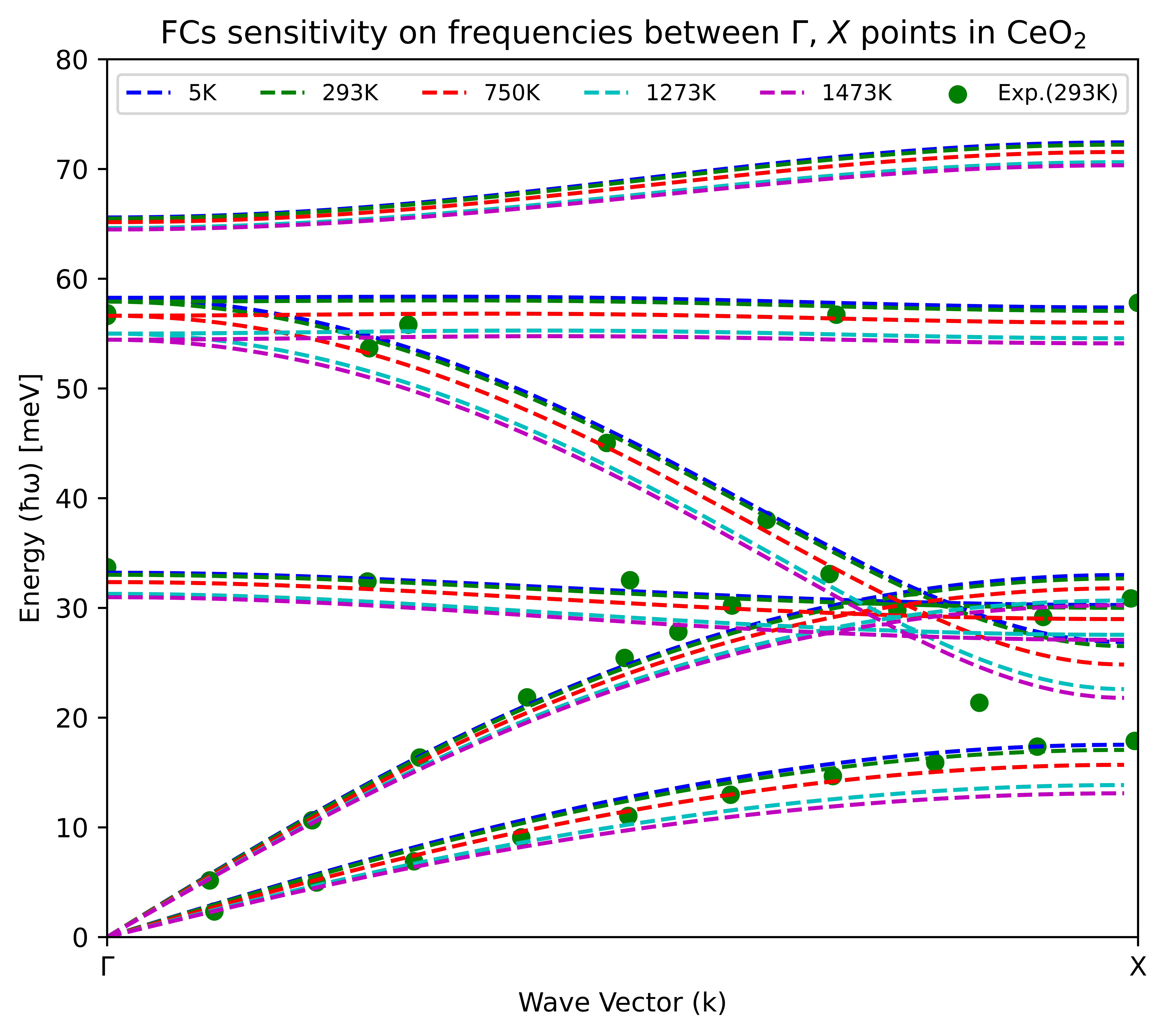}
        \caption{}
        \label{fig:Figure_S4a}
    \end{subfigure}
    \hspace{0.01\textwidth}
    \begin{subfigure}{0.4\textwidth}
        \includegraphics[width=\textwidth]{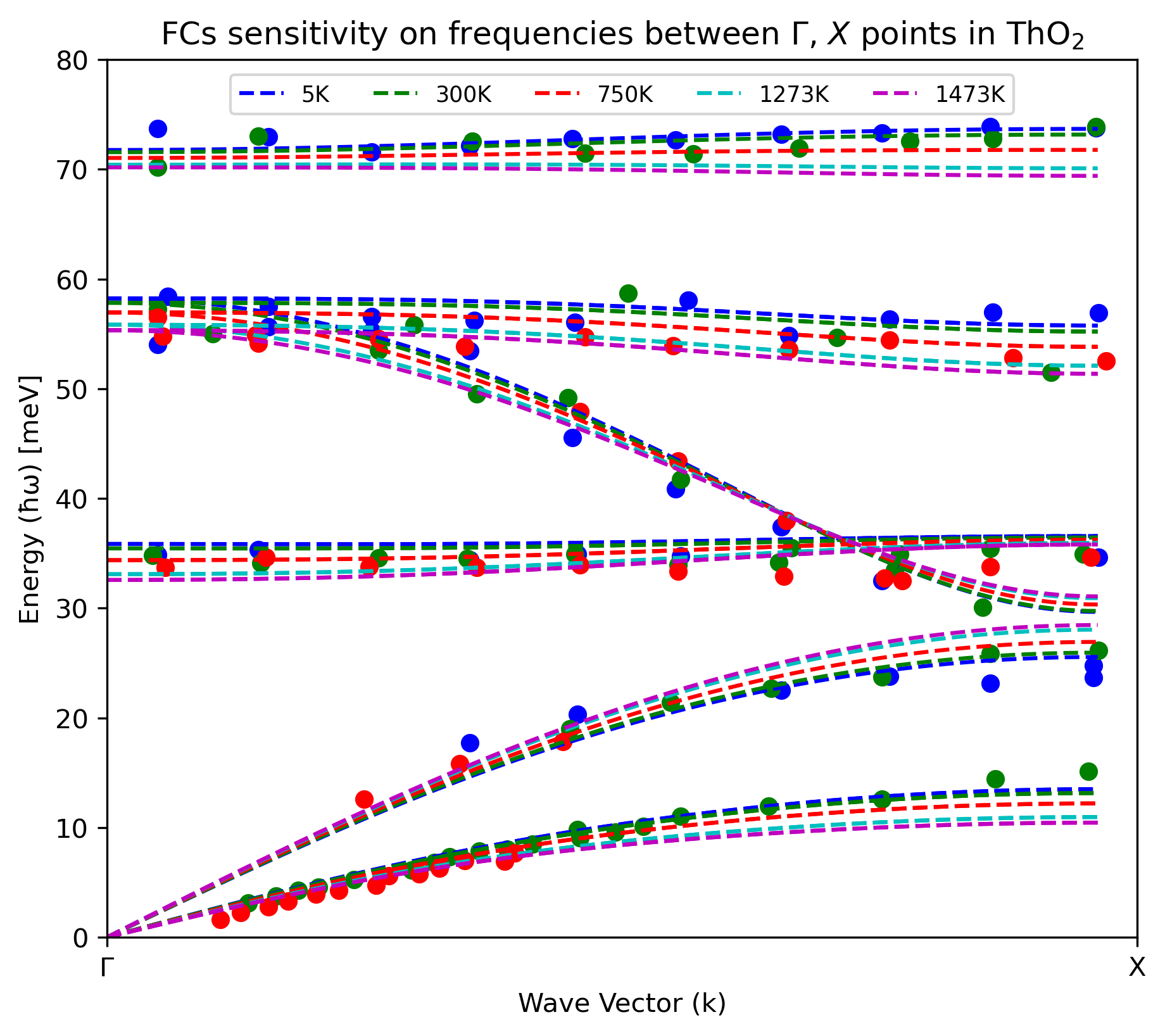}
        \caption{}
        \label{fig:Figure_S4b}
    \end{subfigure}
    \hspace{0.01\textwidth}
    \caption{Temperature-dependence of phonon dispersion (a) CeO$_{2}$ and (b)ThO$_{2}$}
    \label{fig:Figure_S4}
\end{figure*}
\subsubsection{Empirical model with phonon renormalization (EPR)}
This approach accounts for the temperature dependence of the 2nd order interatomic force constants (IFCs) and, for simplicity, considers only 3-phonon interactions, neglecting changes in the 3rd order IFCs. The 2nd order IFCs of thoria were fitted to previously reported temperature-dependent phonon dispersion curves by treating phonon interactions traditionally considered by rigid ion models $F_{\alpha\beta}$ up to the 4th nearest neighbor (Th-O: $\frac{1}{4}\frac{1}{4}\frac{1}{4}$, Th-Th: $0\frac{1}{2}\frac{1}{2}$, O-O: $\frac{1}{2}00$, and O-O: $0\frac{1}{2}\frac{1}{2}$). The phonon dispersion curves used for these fits were obtained from measurements performed on thoria at 5 K, 300 K, and 750 K using inelastic neutron scattering. The 2nd order IFCs in thoria were obtained as a function of temperature by interpolating the reported experimental phonon dispersion data using the Lakkad model expression (based on the Debye model of phonon dispersion), defined as:
\[
F_{\alpha\beta} = F_{\alpha\beta}^0 \left[1 - K_{\alpha\beta} F\left(\frac{T}{\theta_D}\right)\right]
\]

where

\[
F\left(\frac{T}{\theta_D}\right) = 3\left(\frac{T}{\theta_D}\right)^4 \int_0^{\frac{\theta_D}{T}} \frac{x^3}{\exp(x) - 1} \, dx
\]

$\theta_D$ is the Debye temperature, and $K_{\alpha\beta}$ is the anharmonicity parameter for each IFC. The 3rd order IFCs of thoria, considering only 3-phonon interactions, were obtained at ground state (0 K) from previously reported first principles calculations. The phonon linewidth was then extracted as a function of temperature using the Phono3py package with a q-point mesh of 20 $\times$ 20 $\times$ 20. The resulting linewidth calculated using this approach for thoria is shown in Figure \ref{fig:Figure_S2}.

For the case of ceria, since only room temperature phonon dispersion data are available, the parameters in the Lakkad model cannot be uniquely defined. In this instance, a set of parameters that could reproduce the room temperature dispersion was chosen. The temperature-dependent phonon linewidth of ceria was then calculated using the 2nd order IFCs obtained using this approach. The solid lines in \ref{fig:Figure_S3} show that the Raman linewidth exhibits more structure in ceria (when compared to EPR model results for thoria) and displays evidence of linewidth narrowing. This analysis provides a qualitative explanation for a potential mechanism responsible for the observed anomaly. However, it is important to note that the accurate determination of phonon renormalization due to 3- and 4-phonon processes is nontrivial and remains a challenge.

Additional fitting terms from the EPR model are tabulated below. 
\begin{table}[ht]
    \centering
    \caption*{Table S1: Thermally correlated force constants of ThO$_2$ and CeO$_2$ at 0K}
    \begin{tabular}{|c|c|c|c|c|c|}
    \hline
     & $\alpha_{1}$ & $\alpha_{2}$ & $\beta_{1}$ & $\alpha_{3}$ & $\beta_{3}$ \\
    \hline
    ThO$_2$ & -0.0281602 & 0.0630 & -0.0712 & -0.05115 & 0.01135 \\
    \hline
    CeO$_2$ & -0.0219 & 0.0586 & -0.0764 & -0.0581 & 0.0100 \\
    \hline
    \end{tabular}
    \label{tab:supp_table1}
\end{table}

\begin{table}[ht]
    \centering
    \caption*{Table S2: Lakkad’s constant, \(K\) for the modified force constants of ThO$_2$ and CeO$_2$}
    \begin{tabular}{|c|c|c|c|c|c|}
    \hline
     & $\alpha_1$ & $\alpha_2$ & $\beta_2$ & $\alpha_3$ & $\beta_3$ \\
    \hline
    $K$ for ThO$_2$ & 0.0431 & -0.1987 & -0.1177 & 0.0511 & 0.080 \\
    \hline
    $K$ for CeO$_2$ & 0.035 & -0.0322 & 0.0529 & 0.020 & -0.0430 \\
    \hline
    \end{tabular}
    \label{tab:supp_table2}
\end{table}

\begin{figure}[ht]
    \centering
    \includegraphics[width=0.4\textwidth]{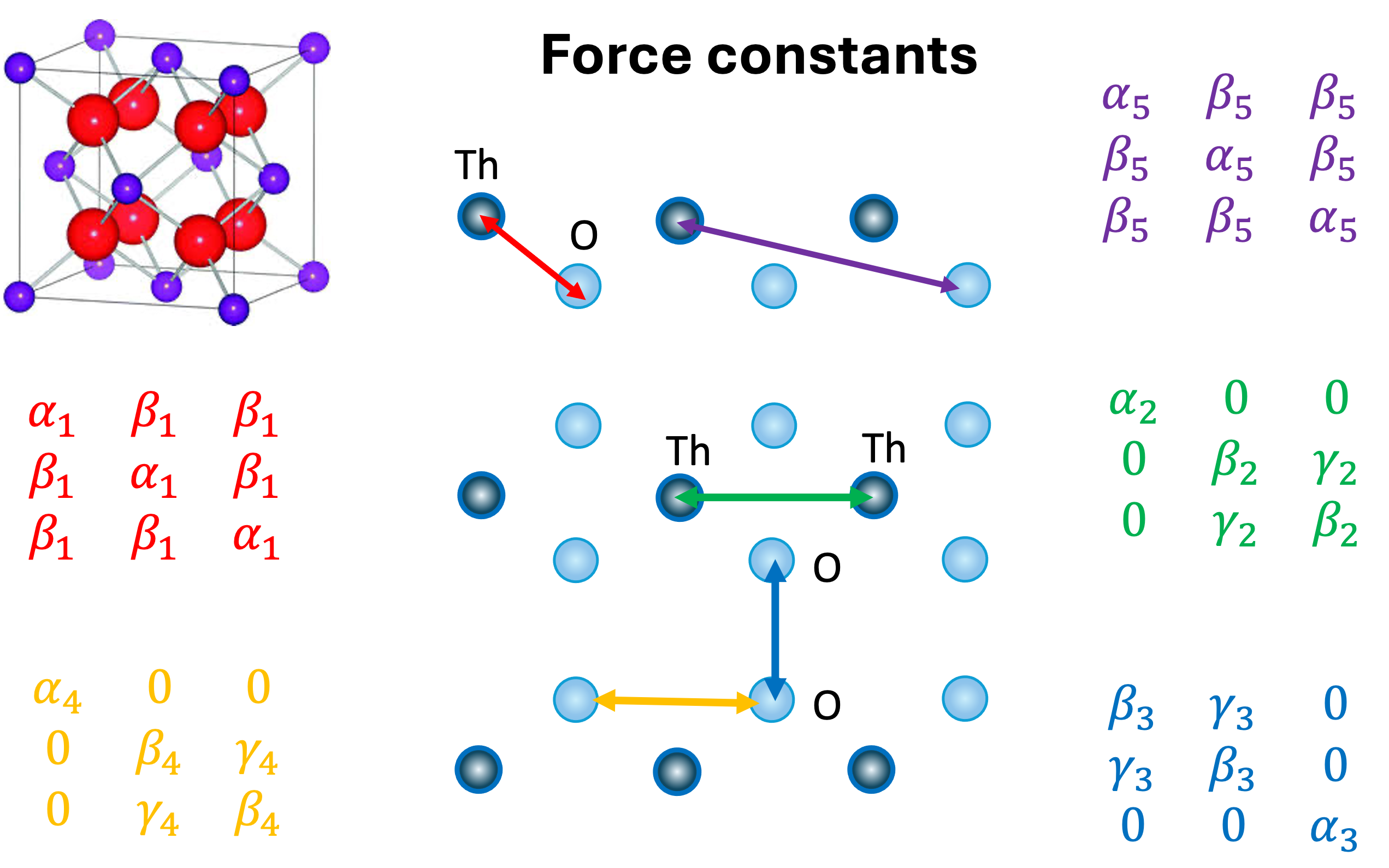}
    \caption{Notation for IFCs used in the EPR model was taken from the Ref.\cite{Lacina_PRB_1970}}
    \label{fig:supp_fig6}
\end{figure}

To conduct this study, we initially utilized force constants computed via Quantum ESPRESSO in Phonopy to plot the dispersion curve employing the rigid ion model for both ThO\(_2\) and CeO\(_2\). Subsequently, we adjusted the previously mentioned force constants according to Lakkad’s expression to account for the thermal effects observed in ThO\(_2\)’s dispersion curve by Adnan et al.\cite{Adnan2023} and to incorporate the thermal effect observed in CeO\(_2\)’s dispersion curve by Clausen et al.\cite{clausen1987inelastic}. Figures S1 and S4 display the dispersion curve of CeO\(_2\) and ThO\(_2\), respectively. Subsequently, we employed Phono3py to observe the T\(_{2g}\)’s Raman peak positions and the T\(_{2g}\)’s Raman linewidth of 24x24 supercell and 20x20x20 meshed CeO\(_2\) and ThO\(_2\) for different temperatures, comparing them with the experimental dataset. These observations were depicted in Figures S2 and S3 for CeO\(_2\) and Figures S5 and S6 for ThO\(_2\). SI\_Figures 1 to 6 strongly elucidate that the anharmonicity of ThO\(_2\) and CeO\(_2\) increases at higher temperatures. Notably, Figure S3 and Figure S6 indicate that CeO\(_2\) exhibits a higher T\(_{2g}\) Raman linewidth than ThO\(_2\), suggesting that thermal effects are more pronounced in CeO\(_2\). This higher anharmonicity reduces CeO\(_2\)’s phonon lifetime and thermal conductivity compared to ThO\(_2\) at the same elevated temperature.
\section{References}
\bibliography{Ref}

\end{document}